\documentclass[twocolumn]{svjour3}

\usepackage{amsmath}
\usepackage{amssymb}
\usepackage{amsfonts}
\usepackage{mathptmx}
\usepackage{graphicx}
\usepackage{epstopdf}
\usepackage{dcolumn}
\usepackage{bm}

\DeclareMathOperator{\tr}{\mathop{\mathrm{Tr}}}

\DeclareMathOperator{\re}{\mathop{\mathrm{Re}}}

\newcommand{\Eq}[1]{Eq.~(\ref{#1})}

\begin{document}

\title{Andreev current and subgap conductance of spin-valve SFF structures}

\author{A.~S.~Vasenko \and A.~Ozaeta \and S.~Kawabata \and F.~W.~J.~Hekking \and F.~S.~Bergeret}

\institute{A.~S.~Vasenko \at
              Institut Laue-Langevin, 6 rue Jules Horowitz, BP 156,\\
              38042 Grenoble, France\\
              \email{vasenko@ill.fr}
           \and
              A.~Ozaeta \and F.~S.~Bergeret \at
              Centro de F\'{i}sica de Materiales (CFM-MPC),\\ Centro
              Mixto CSIC-UPV/EHU, Manuel de Lardizabal 5,\\ E-20018 San
              Sebasti\'{a}n, Spain;\\
              Donostia International Physics Center (DIPC),\\ Manuel
              de Lardizabal 4, E-20018 San Sebasti\'{a}n, Spain
           \and
               S.~Kawabata \at
               Electronics and Photonics Research Institute (ESPRIT),\\ National Institute of Advanced Industrial Science\\
               and Technology (AIST),\\ Umezono 1-1-1, Tsukuba, Ibaraki 305-8568, Japan
           \and
               F.~W.~J.~Hekking \at
               NLPMMC, Universit\'{e} Joseph Fourier and CNRS,\\ 25
               Avenue des Martyrs, BP 166, 38042 Grenoble, France
}

\date{Received: date / Accepted: date}

\maketitle

\begin{abstract}
The Andreev current and the subgap conductance in a superconductor/ insulator/ ferromagnet (SIF) structure
in the presence of a small spin-splitting field show novel interesting features \cite{Ozaeta1}. For example, the Andreev
current at zero temperature can be enhanced by a spin-splitting field $h$, smaller than the superconducting gap $\Delta$,
as has been recently reported by the authors. Also at finite temperatures the Andreev current has a peak for values
of the spin-splitting field close to the superconducting gap, $h \approx \Delta$. Finally, the differential subgap conductance
at low temperatures show a peak at the bias voltage $eV = h$. In this paper we investigate
the Andreev current and the subgap conductance in SFF structures with arbitrary direction of magnetization of the F layers.
We show that all
aforementioned features occur now at the value of the ``effective field'', which is the field acting on the Cooper pairs
in the multi-domain ferromagnetic region, averaged over the decay length of the superconducting condensate into a ferromagnet.
We also briefly discuss the heat transport and electron cooling in the considered structures.
\end{abstract}


\section{Introduction}

Ferromagnetism and superconductivity are antagonistic to each other's orders, however their interplay can be realized
when the two interactions are spatially separated. In this case the coexistence of the two orderings is due to the proximity
effect \cite{RevG}, \cite{RevB}, \cite{RevV}. Experimentally this situation can be
realized in superconductor/ ferromagnet (S/F) hybrid structures. The role of
the Andreev reflection is central to the proximity effect since it provides the mechanism for converting single electron
states from a normal (N) or ferromagnetic metal to Cooper pairs in the superconducting condensate \cite{Andreev}, \cite{Pannetier}.
During the Andreev reflection
process the electron incoming to the N/S (F/S) interface is reflected as a hole and a charge $2e$ is transferred across the interface.
As a result a long-range electron-hole coherence is induced into the non-superconducting material. The Andreev reflection
manifests itself in the subgap conductance, i.e. the conductance for voltages smaller than the superconducting gap $\Delta$.
In diffusive N/S systems the subgap conductance shows the zero bias anomaly peak due to the impurity confinement and the electron-hole
interference at the Fermi level \cite{Kastalsky}, \cite{Klapwijk}, \cite{Hekking}.

At a S/F interface the mechanism of Andreev reflection is modified compared to the N/S hybrid
structures since the incoming electron and reflected hole belong to different spin bands \cite{deJong}.
Thus, one expects a suppression of the
Andreev (subgap) current by increasing the exchange field $h$, which is a measure of the spin-splitting at the Fermi level.
Recently it was shown by the authors that this intuitive picture does not hold always \cite{Ozaeta1}. If the voltage exceeds some
critical value the Andreev current of a tunnel ferromagnet/ insulator/ superconductor (FIS)
structure is enhanced by a small exchange field $h < \Delta$
reaching a maximum at $h \approx eV$ at zero temperature. If one keeps the voltage low but now increase the temperature,
the Andreev current (as well as the full current at this temperature)
shows a peak at $h \approx \Delta$. All these novel features
were exhaustively discussed in \cite{Ozaeta1}. Finally it was shown that the subgap conductance of a FIS junction at low
temperatures and small exchange fields $h < \Delta$ has a peak at $eV = h$ \cite{Ozaeta1}, \cite{Leadbetaer1999}.
Thus its measurement can be used to determine the
strength of a weak exchange or Zeeman-like field in the hybrid structure. The latter can be not only the intrinsic exchange
field of a ferromagnetic alloy \cite{small_h} but also a spin-splitting field created in the normal metal by a magnetic induction $B$
(in which case $h = \mu_B B$, where $\mu_B$ is the Bohr magneton) or by a proximity to the
ferromagnetic insulator material \cite{Cottet2011}.

All these predictions were made in \cite{Ozaeta1} for the mono-domain FIS hybrid system. The purpose of this work is
to consider hybrid structure with a multi-domain ferromagnetic metal. We present a quantitative analysis of the electron
transport in FIS tunnel structures where a ferromagnetic layer consist of two magnetic domains with arbitrary direction of
magnetization (so called ``superconducting spin-valve'' \cite{Karminskaya}). We show that in this case the aforementioned features
of the Andreev current and subgap conductance occur at the value of the ``effective field'', which is the field acting on the Cooper pairs
in the multi-domain ferromagnetic region, averaged over the decay length of the superconducting condensate into a ferromagnet \cite{Bergeret_h}.


\section{Model and basic equations}

\begin{figure}[t]\begin{center}
\includegraphics[scale=0.33]{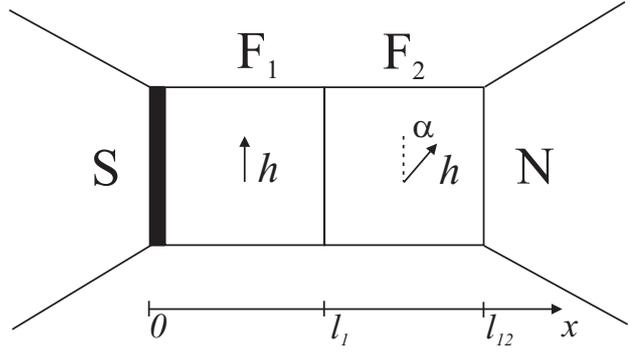}
\caption{The  SF$_1$F$_2$N junction.
The interface at $x=0$ corresponds to the insulating barrier (thick black line). Interfaces at $x=l_1$ and $x=l_{12}$ are fully transparent.
 $\alpha$ is the angle between the magnetization directions of  F$_1$ and F$_2$.} \label{model} \vspace{-4mm}
\end{center}
\end{figure}

The model of a SF$_1$F$_2$N junction we are going to study is depicted in Fig.~\ref{model}
and consists of a ferromagnetic bilayer F$_1$F$_2$ of thickness $l_{12}=l_1+l_2$ connected
to a superconductor (S) and a normal (N) reservoirs along the $x$ direction.
We consider the diffusive limit, i.e the elastic scattering length $\ell$ is much smaller than
the decay  length of the superconducting condensate into a ferromagnet $\xi_h = \sqrt{\mathcal{D}/ 2 h}$
and the superconducting coherence length $\xi = \sqrt{\mathcal{D}/ 2 \Delta}$,
where $\mathcal{D}$ is the diffusion coefficient and $h$ is the value of the exchange field
(we set $\hbar =k_{B}=1$ and for simplicity we assume the same $\mathcal{D}$ in the whole structure).
We also assume that the F$_1$F$_2$ and F$_2$N interfaces are transparent, while the SF$_1$ is a tunnel barrier.
Thus, the two ferromagnetic layers are kept at the same potential as the voltage-biased normal reservoir.
The F$_1$F$_2$ bilayer can either model a two domain ferromagnet or an artificial hybrid magnetic structure.

The magnetization of the F$_1$ layer is along the $z$ direction, while the magnetization of the
F$_2$ layer forms an angle $\alpha$ with the  one of the layer F$_1$. Both magnetization vectors lie in the $yz$ plane.
Correspondingly the exchange field vector in the F$_1$ is given by
${\bf h} = (0, 0, h)$, and in the F$_2$ layer
by ${\bf h} = (0, h\sin\alpha, h\cos\alpha)$, where
the angle $\alpha$ takes values from  0 (parallel configuration)
to  $\pi$ (antiparallel configuration).

Under these conditions, the microscopic calculation of the electric current
through the structure requires solution of the quasiclassical equation
for the $8\times 8$ Keldysh-Green function $\breve{G}$ in the Keldysh $\times$ Nambu $\times$ spin space
in the F$_1$F$_2$ bilayer \cite{Usadel}, \cite{Ivanov},
\begin{equation}\label{Usadel3D}
i\mathcal{D} \partial_x \breve{J} = \left[ \breve{H},
\breve{G}\right], \quad \breve{G}^2 = 1, \quad \breve{G} = \begin{pmatrix} \check{G}^R & \check{G}^K \\
0 & \check{G}^A
\end{pmatrix}.
\end{equation}
Here $\breve{H} = \tau_z\left(E  - {\bf h} {\bf \sigma}\right)$ is the Hamiltonian, $\breve{J} = \breve{G}\partial_x \breve{G}$ is the
matrix spectral current,
${\bf \sigma} = (\sigma_x, \sigma_y, \sigma_z)$ are the Pauli matrices in spin space and $\tau_z$  in Nambu space.
The $R$, $A$ and $K$ indices stand for the retarded, advanced and Keldysh components (we use the symbols $\breve .$  for $8\times 8$
and $\check .$  for $4\times 4$ matrices). In \Eq{Usadel3D} we neglect the inelastic collision term, assuming $l_{12}$ to be
smaller than the inelastic relaxation length \cite{Arutyunov}.

In the F$_1$ region  ${\bf h \sigma} = h \sigma_z$ and the equation \Eq{Usadel3D} has the form
\begin{equation}\label{Usadel}
i\mathcal{D} \partial_x \breve{J} = \left[ \tau_z\left(E  - \sigma_z h\right),
\breve{G}\right].
\end{equation}
In the F$_2$ region ${\bf h \sigma} = h \sigma_z \exp(-i \sigma_x \alpha)$ and
it is convenient to introduce Green's functions rotated in spin space
\cite{Bergeret2002},
\begin{equation}\label{gauge}
\widetilde{\breve{G}} = U^\dagger \breve{G} U, \quad U =
\exp\left( i \sigma_x \alpha/2 \right).
\end{equation}
The  rotated function $\widetilde{\breve{G}}$ is then determined by \Eq{Usadel}.

The \Eq{Usadel} should be complemented by boundary conditions at the interfaces.
As mentioned above, we assume that
the F$_1$F$_2$ and F$_2$N interfaces are transparent and therefore the boundary conditions at $x=\l_1,l_{12}$ read
\begin{eqnarray}
\breve{G} \bigl |_{x=l_1 - 0}& =& \breve{G} \bigl |_{x=l_1 + 0}\label{bc1},\\
\partial_x \breve{G} \bigl |_{x=l_1 - 0}& =& \partial_x \breve{G}
\bigl |_{x=l_1 + 0}\label{bc11},\\
\breve{G}\bigl|_{x=l_{12}-0} &=&\tau_z.
\end{eqnarray}
At $x=0$,  the SF$_1$ interface is a tunnel barrier, where the boundary
conditions are given by the relation \cite{KL},
\begin{equation}\label{kupluk}
\breve{J} \bigl |_{x=0} = (W/\xi) \left[
\breve{G}_S, \breve{G} \right]_{x=0}.
\end{equation}
Here $\breve G_S$ is the Green function of a bulk BCS superconductor defined as
\begin{subequations}
\begin{align}
\breve{G}_S &= \tau_z u + \tau_x v, \label{super}
\\
(u, v) &= (E, \; i\Delta)/\epsilon, \quad \epsilon =
\sqrt{(E+i\eta)^2 - \Delta^2}, \label{uv}
\end{align}
\end{subequations}
$W\ll 1$ is  the diffusive transparency parameter \cite{Chalmers},
$W = \xi/ 2 g_N R$, and $\eta$ is the Dynes parameter \cite{Dynes}.
In our calculations we set small $\eta = 10^{-3} \Delta_0$ where
$\Delta_0$ is the superconducting gap at zero temperature. Below we omit
$\eta$ in analytical expressions for simplicity.

The electric current through the structure is given by the following expression \cite{LOnoneq},
\cite{Belzig},
\begin{equation}\label{I}
I = \frac{g_N}{8 e} \int_0^{\infty} \tr \tau_z \check{J}^K \; dE,
\end{equation}
where  $\breve{J}^K \equiv \left( \breve{G}\partial_x \breve{G} \right)^K =
\check{G}^R \partial_x \check{G}^K + \check{G}^K \partial_x \check{G}^A$.
By neglecting non-equilibrium effects,
the Keldysh component of Green's function is related to the retarded and advanced ones by
\begin{subequations}
\begin{align}
\check{G}^K &= \check{G}^R \check{n} - \check{n} \check{G}^A,
\quad \check{n} = n_+ + \tau_z n_-,
\\
n_\pm &= \frac{1}{2}\left( \tanh \frac{E + eV}{2T} \pm
\tanh\frac{E - eV}{2T} \right),
\end{align}
\end{subequations}
where $n_\pm$ and $T$ are correspondingly the equilibrium quasiparticle distribution functions and the temperature.
Below we express the advanced Green functions through the retarded ones using the general relation
$\check{G}^A =-\tau_z \check{G}^{R \dagger} \tau_z$ \cite{LOnoneq}.

In particular, we are interested in the Andreev current, i.e. the current for voltages smaller than the superconducting
gap due to Andreev processes at the SF$_1$ interface. It is given by the expression \cite{VZK}, \cite{VBCH},
\begin{equation}\label{IA}
I_A = \frac{1}{eR} \int_0^\Delta n_-(E) M_S(E) \re f_0 \; dE.
\end{equation}
where $M_S(E) = \Delta \Theta(\Delta - |E|)/\sqrt{\Delta^2 - E^2}$ is the condensate spectral function,
$\Theta(x)$ is the Heaviside step function and the function $f_0$ is the singlet component of $\hat{f}$ at $x = 0$.
This equation is used throughout the article to determine the Andreev transport. We neglect the contribution
to the Andreev current due to the partial Andreev reflection at the energies above the superconducting gap.
In the case of strong enough tunnel barrier at $x=0$ this contribution leads to negligible corrections \cite{VBCH}.

Because of the  low transparency of the tunnel SF$_1$
barrier, the proximity effect is weak and the  retarded Green function can be linearized (we omit the superscript $R$),
\begin{equation}\label{G_lin}
\check{G} \approx \tau_z + \tau_x \hat{f},
\end{equation}
where $\hat{f}$ is the $2 \times 2$ anomalous Green function in the spin space ($|\hat{f}|\ll 1$) that obeys  the linearized equation,
\begin{equation}\label{linearized}
i\mathcal{D} \partial^2_{xx} \hat{f} = 2 E \hat{f} - \left\{ {\bf h
\sigma}, \hat{f} \right \},
\end{equation}
where $\{\cdot, \cdot\}$ stands for the anticommutator.
The general solution of this equation has the form
\begin{equation}\label{f}
\hat{f}(x) = f(x) + f_y(x) \sigma_y + f_z(x) \sigma_z,
\end{equation}
where $f$ is the singlet component and $f_{z, y}$ are the triplet components with respectively
zero and $\pm 1$ projections on the spin quantization axis \cite{Bergeret}.

Solving \Eq{linearized} in the F$_1$ layer we obtain for the components of \Eq{f},
\begin{subequations}\label{f_i}
\begin{align}
f_\pm(x) &= a_\pm \cosh (k_\pm x) + \frac{2W}{k_\pm} (u a_\pm - v)
\sinh (k_\pm x),
\\
f_y(x) &= a_y \cosh (k_y x) + \frac{2W}{k_y} u a_y \sinh (k_y x),
\end{align}
\end{subequations}
where $f_\pm = f \pm f_z$,  $a_i$ are the boundary values of $f_i$ at  $x=0$
($i$ stands for $+,-,y$) and the characteristic wave vectors are
\begin{equation}
k_\pm = \sqrt{\frac{2(E \mp h)}{i\mathcal{D}}}, \quad
k_y =\sqrt{\frac{2E}{i\mathcal{D}}}.
\end{equation}
In the F$_2$ layer  the general solution has the form,
\begin{equation}
\widetilde{f}_i(x) = b_i \sinh \left[ k_i (x - l_{12}) \right],
\end{equation}
where $\widetilde{f}_i$ are the components of the rotated Green function, \Eq{gauge}.
Using the boundary conditions at the F$_1$F$_2$ interface, Eqs. (\ref{bc1}-\ref{bc11}) we obtain a set of six linear equations for the
six coefficients $a_i$ and $b_i$, that can be solved straightforwardly.
In particular we are interested in $f_0=(a_++a_-)/2$ which enters the equation for the Andreev current, \Eq{IA}.
Since the analytical expression is cumbersome we do not present it here.


\section{Results and discussion}

\begin{figure*}[t]\begin{center}
\includegraphics[scale=0.40]{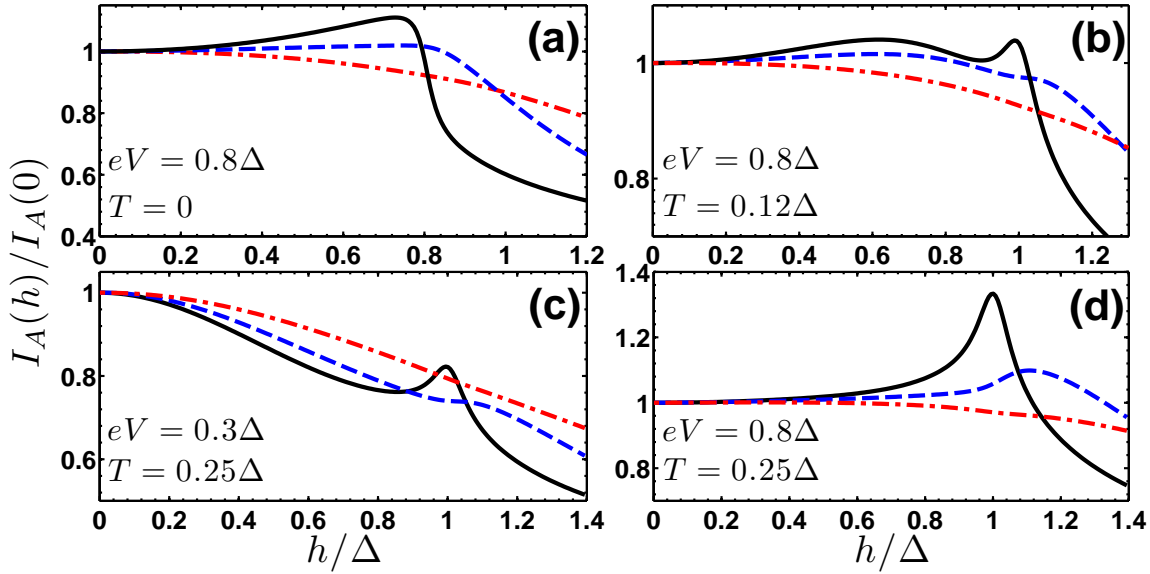}
\vspace{4mm}
\caption{The $h$-dependence of the ratio $I_A(h)/I_A(0)$ for $l_1 = \xi$ and $l_2 = 9\xi$, $W = 0.007$, $\alpha = 0$ (solid
black line), $\alpha = \pi/2$ (dashed blue line) and $\alpha = \pi$ (dash-dotted red line). (a) $eV = 0.8 \Delta$, $T = 0$;
(b) $eV = 0.8 \Delta$, $T = 0.12 \Delta$; (c) $eV = 0.3 \Delta$, $T = 0.25 \Delta$; (d) $eV = 0.8 \Delta$, $T = 0.25 \Delta$.}
\label{Fig1} \vspace{-4mm}
\end{center}
\end{figure*}

First we briefly review the novel features of the Andreev current for a mono-domain SIFN structure ($\alpha = 0$),
discussed in \cite{Ozaeta1}.
At zero temperature we observe the enhancement of the Andreev current at high enough voltages above some critical value,
see Fig.~\ref{Fig1} (a), solid black line. The Andreev current first increases by increasing $h$, reaches a maximum at
$h \approx eV$, and then decays by further increase of the exchange field.
The enhancement of the Andreev current is due to the competition between two-particle tunneling processes
and decoherence mechanisms.
Sharp suppression of the Andreev current at $h \approx eV$ occurs when the electron-hole coherence length
$\sqrt{\mathcal{D}/ 2 eV}$ is cut off by the
decay length of superconducting condensate into a ferromagnet, $\xi_h = \sqrt{\mathcal{D}/ 2 h}$.

Another feature of the Andreev current, predicted in \cite{Ozaeta1} is the peak at $h \approx \Delta$ which can
be observed only at finite
temperatures. The relative height of this peak increases with temperature and voltage,
see Figs.~\ref{Fig1} (c) and (d), solid black line. In case of
large enough values of $V$ and $T$, we observe both the enhancement of the Andreev current
by increasing $h$ and the peak at $h \approx \Delta$, see
Fig.~\ref{Fig1} (b), solid black line. The peak can be observed only for high enough
temperatures when the upper limit of the integration in \Eq{IA}
is $\Delta$ (at zero temperature the upper limit is $eV < \Delta$).
Then the integrand in \Eq{IA} has a ``dangerous point'' at $E = h = \Delta$.
This peak can be observed by measuring the full electric current through the
junction as the single particle current is almost independent on $h$.
Note that for the values of temperature used in our calculations
$\Delta \approx \Delta_0$.

\begin{figure*}[t]\begin{center}
\includegraphics[scale=0.35]{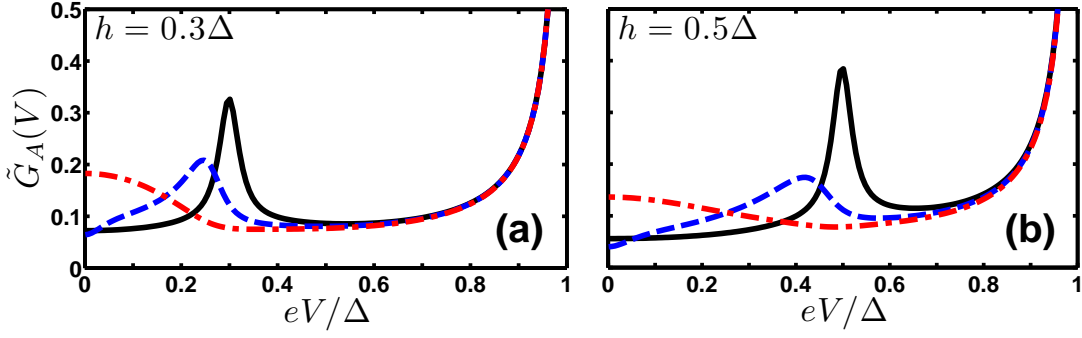}
\vspace{4mm}
\caption{The bias voltage dependence of the differential conductance at $T=0$ for exchange
fields (a) $h = 0.3 \Delta$ and (b) $h = 0.5 \Delta$ for $l_1 = \xi$ and $l_2 = 9\xi$, $W = 0.007$, $\alpha = 0$ (solid
black line), $\alpha = \pi/2$ (dashed blue line) and $\alpha = \pi$ (dash-dotted red line).
Here $\tilde{G}_A= 4 R_T G_A$.} \label{Fig2} \vspace{-4mm}
\end{center}
\end{figure*}

Now let us reconsider these features for the two-domain situation in case of $\alpha = \pi/2$
(dashed blue lines in Fig.~\ref{Fig1}) and $\alpha = \pi$
(dash-dotted red line in Fig.~\ref{Fig1}). The thickness of the F layers is chosen to be
$l_1 = \xi$ and $l_2 = 9\xi$, $l_1$ short enough for the
superconducting condensate penetrates both ferromagnetic layers and $l_2$ long enough for
the full development of the proximity effect in F$_1$F$_2$ bilayer
(at small values of $l_2$ the Andreev current is suppressed by the proximity of the normal
reservoir at $x = l_{12}$) \cite{Ozaeta2}.

Firs of all, we see that increasing $\alpha$ the features (peaks at $h \approx eV, \Delta$)
smear and their amplitude reduces. For $\alpha = \pi$ we do not see any more the enhancement of
the Andreev current. Secondly, we see shift of these peaks to the larger
values of $h$, which is explicitly seen for $\alpha = \pi/2$. The peak at $h \approx eV$ is
shifted to the right (Fig.~\ref{Fig1} (a), dashed blue line) as
well as the peak at $h \approx \Delta$ (Fig.~\ref{Fig1} (d), dashed blue line).
This can be explained as follows. The superconducting condensate penetrates
both ferromagnetic layers and feel the ``effective exchange field'' $\bar h$
acting on the Cooper pairs, averaged over the length $\xi_h$ \cite{Bergeret_h}.
The $\bar{h}(\alpha)$ is gradually reduced as $\alpha$ increases from $0$ to $\pi$.
As before the Andreev current peak is at $\bar{h}(\alpha) \approx \Delta$
which in the case of a finite $\alpha$ corresponds to larger values of the bare $h$, therefore
we observe shift of the Andreev current peak to the right.

\begin{figure*}[t]\begin{center}
\includegraphics[scale=0.35]{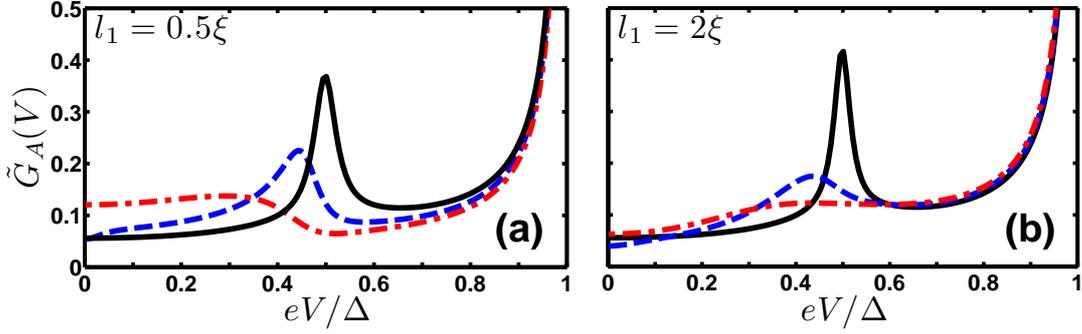}
\vspace{4mm}
\caption{The bias voltage dependence of the differential conductance at $T=0$ for exchange
field $h = 0.5 \Delta$ for (a) $l_1 = 0.5 \xi$ and (b) $l_1 = 2\xi$, $l_2 = 9\xi$, $W = 0.007$, $\alpha = 0$ (solid
black line), $\alpha = \pi/2$ (dashed blue line) and $\alpha = \pi$ (dash-dotted red line).
Here $\tilde{G}_A= 4 R_T G_A$.} \label{Fig3}
\vspace{-4mm}
\end{center}
\end{figure*}

Let us now calculate the subgap differential conductance $G_A = dI_A/dV$ at zero temperature.
It is known that for a diffusive NIS junction the differential conductance at low temperatures has a
peak at $eV = \Delta$ and a zero bias
anomaly (ZBA) peak due to the impurity confinement and the electron-hole
interference at the Fermi level \cite{Kastalsky}, \cite{Klapwijk}, \cite{Hekking}.
It occurs at zero bias since for $V = 0$ the electron is perfectly retro-reflected as a hole
during the Andreev reflection process. Thus the electron and the reflected hole
interfere along the same trajectory and the interference effect strongly enhance the
subgap conductance at zero bias \cite{Klapwijk}.

For the FIS structures with $h < \Delta$ the ZBA peak is now shifted to the finite voltage
$eV = h$ \cite{Ozaeta1}, \cite{Leadbetaer1999}, see Fig.~\ref{Fig2}, solid black lines. This can be described as follows. Upon entering
of the Cooper pair into the ferromagnetic metal the
spin up electron in the pair lowers its potential energy by $h$, while the spin down
electron raises its potential energy by the same amount.
In order for each electron to conserve its total energy, the
spin up electron must increase its kinetic energy, while the
spin down electron must decrease its kinetic energy, to make
up for these additional potential energies in F \cite{Demler}. Therefore the electron-hole pair in F has now the
momentum mismatch, i.e. the electron is not perfectly retro-reflected. However, if $eV = h$ there is
a possibility for exact retro-reflection (and interference along the trajectory) of an electron to a hole with
a same kinetic energy equal to the Fermi energy.

In case of the two-domain ferromagnetic metal we now have the ZBA shift to the
``effective exchange field'' $eV = \bar{h}(\alpha)$. The ``effective field'' is smaller
than the bare $h$, $\bar{h}(\alpha) < h$, and therefore we observe the shift of the differential
conductance peak to the left.
We can explicitly see this for $\alpha = \pi/2$
(Fig.~\ref{Fig2}, dashed blue lines). For $\alpha = \pi$ the situation is more complicated as the effective
exchange field is rather small in the antiparallel configuration.
For $l_1 = \xi$ we observe a broad ZBA peak at $V = 0$ for $\alpha = \pi$ for both values of $h = 0.3 \Delta$
and $0.5 \Delta$ (Fig.~\ref{Fig2}, dash-dotted red lines).
For $l_1 \neq \xi$ the maximum is shifted from the zero bias, see Fig.~\ref{Fig3}, dash-dotted red lines.

In the end we notice that SIFFN structures with two-domain ferromagnetic interlayer show interesting
behavior of the heat current through the structure \cite{Ozaeta2}. It is known that in NIS tunnel junctions
the flow of electric current is accompanied by a heat transfer from the normal metal into the superconductor
\cite{Giazotto2006}, \cite{Pekola_A}. This happens due to the selective tunneling of high-energy quasiparticles
out of the normal metal in presence of the superconducting energy gap $\Delta$. The heat transfer through NIS
junctions can be used for the realization of a microcooler, and the important problem is to overcome
possible limitations of its cooling performance. Some of the limitations arise from the fact that nonequilibrium
quasiparticles injected into the superconducting electrode accumulate near the tunneling interface
\cite{VH}, \cite{RCP}. This problem can be solved by imposing a local thermal equilibrium in the superconductor
electrode by means of a ``quasiparticle trap'', i.e. an additional normal metal layer covering the superconductor
electrode \cite{trap1}, \cite{trap2}. Another fundamental limitation arise from the Andreev reflection processes:
the Andreev current $I_A$ does not transfer heat through the N/S interface but rather generates the Joule heating
$I_A V$ which fully dissipates in the normal electrode and dominates quasiparticle cooling at low temperatures
\cite{VBCH}, \cite{Sukumar}. In order to reduce this factor it was proposed to add a ferromagnetic interlayer in
the NIS structure to suppress the Andreev current and enhance the heat current and cooling performance \cite{Giazotto}.

From our studies we can conclude that the ferromagnetic interlayer with small enough exchange field will rather
enhance the Andreev current and suppress the heat current (cooling power) through the structure. One need ferromagnet with an
exchange field (bare or effective in multi-domain case) larger than the superconductor gap $\Delta$ to suppress the
Andreev reflection processes and enhance the cooling performance. The $\alpha$-dependence of the heat current in SIFFN structures
was discussed in \cite{Ozaeta2}.

\section{Conclusions}

To summarize, we have studied the Andreev current and the subgap conductance behavior in SIFFN hybrid structures
with arbitrary direction of magnetization of the F layers. We have revisited all novel features predicted recently in the mono-domain
SIF system in the presence of a small spin-splitting field $h$ \cite{Ozaeta1}, namely the Andreev current peaks at $h \approx eV$
at $T=0$ and at $h \approx \Delta$ for high enough temperature, and the differential conductance peak at
$eV = h$. We have shown that in the two-domain case the aforementioned features occur at the value of
the ``effective exchange field'' $\bar{h}(\alpha) < h$, which is the field acting on the
Cooper pairs in the multi-domain ferromagnetic region, averaged over
the decay length of the superconducting condensate into a ferromagnet, $\xi_h$. Increasing $\alpha$ from $0$ to $\pi$ one gradually
reduce the effective field $\bar{h}(\alpha)$. We also briefly discuss the heat transport and electron cooling in the considered structures.


\begin{acknowledgements}
This work was supported by  the Spanish Ministry of Economy and Competition under
Project FIS2011-28851-C02-02, the Basque Government under UPV/EHU Project IT-366- 07,
the ``Topological Quantum Phenomena'' (No.22103002) KAKENHI on Innovative Areas,
a Grant-in-Aid for Scientific Research (No. 22710096) from MEXT of Japan,
and JSPS Institutional Program for Young Researcher Overseas Visits.
The work of A.O. was supported by the CSIC and the European Social Fund under JAE-Predoc
program.
\end{acknowledgements}


\begin{thebibliography}{99}

\bibitem{Ozaeta1}
A.  Ozaeta, A. S. Vasenko, F. W. J. Hekking  and F. S. Bergeret:
Phys. Rev. B \textbf{86}, 060509(R) (2012)

\bibitem{RevG}
A.~A.~Golubov, M.~Yu.~Kupriyanov, E.~Il'ichev: Rev.\ Mod.\
Phys.\ \textbf{76}, 411 (2004)

\bibitem{RevB}
A.~I.~Buzdin: Rev.\ Mod.\ Phys.\ \textbf{77}, 935 (2005)

\bibitem{RevV}
F.~S.~Bergeret, A.~F.~Volkov, K.~B.~Efetov: Rev.\ Mod.\
Phys.\ \textbf{77}, 1321 (2005)

\bibitem{Andreev}
A.~F.~Andreev: Zh. Eksp. Teor. Fiz. {\bf 46}, 1823 (1964); [Sov. Phys. JETP
{\bf 19}, 1228 (1964)]; D.~Saint-James: J. Phys. (Paris) {\bf 25}, 899 (1964)

\bibitem{Pannetier}
B.~Pannetier, H.~Courtois: J. of Low Temp. Phys. {\bf 118}, 599 (2000)

\bibitem{Kastalsky}
A.~Kastalsky, A.~W.~Kleinsasser, L.~H.~Greene, R.~Bhat, F.~P.~Milliken, and J.~P.~Harbison: Phys. Rev. Lett. {\bf 67}, 3026 (1991)

\bibitem{Klapwijk}
B.~J.~van~Wees, P.~de~Vries, P.~Magn\'{e}e, and T.~M.~Klapwijk: Phys. Rev. Lett. {\bf 69}, 510 (1992)

\bibitem{Hekking}
F.~W.~J.~Hekking and Yu.~V.~Nazarov: Phys. Rev. Lett. {\bf 71}, 1625 (1993);
Phys. Rev. B {\bf 49}, 6847 (1994)

\bibitem{deJong}
M.~J.~M.~de Jong and C.~W.~J.~Beenakker: Phys. Rev. Lett. {\bf 74},
1657 (1995)

\bibitem{Leadbetaer1999}
M.~Leadbeater, C.~J.~Lambert, K.~E.~Nagaev, R.~Raimondi, and
A.~F.~Volkov: Phys. Rev. B {\bf 59}, 12264 (1999)

\bibitem{small_h}
T.~Kontos, M.~Aprili, J.~Lesueur, X.~Grison, and L.~Dumoulin:
Phys.\ Rev.\ Lett.\, \textbf{93}, 137001 (2004)

\bibitem{Cottet2011}
A.~Cottet: Phys. Rev. Lett. {\bf  107}, 177001 (2011)

\bibitem{Karminskaya}
T. Yu. Karminskaya, A. A. Golubov, and M. Yu. Kupriyanov: Phys. Rev. B \textbf{84}, 064531 (2011)

\bibitem{Bergeret_h}
F. S. Bergeret, A. F. Volkov, and K. B. Efetov: Phys. Rev. Lett. \textbf{86},
3140 (2001)

\bibitem{Usadel}
K.~D.~Usadel: Phys.~Rev.~Lett. {\bf 25}, 507 (1970)

\bibitem{Ivanov}
D.~A.~Ivanov and Ya.~V.~Fominov: Phys. Rev. B \textbf{73} 214524
(2006)

\bibitem{Arutyunov}
K.~Yu.~Arutyunov, H.-P.~Auraneva, and A.~S.~Vasenko: Phys. Rev. B {\bf 83}, 104509 (2011)

\bibitem{Bergeret2002}
F.S. Bergeret, A. F. Volkov and K. B. Efetov: Phys. Rev. B \textbf{66}, 184403 (2002)

\bibitem{KL}
M.~Yu.~Kuprianov and V.~F.~Lukichev: Zh. Eksp. Teor. Fiz. {\bf
94}, 139 (1988); [Sov. Phys. JETP {\bf 67}, 1163 (1988)]

\bibitem{Chalmers}
E.~V.~Bezuglyi, A.~S.~Vasenko, V.~S.~Shumeiko, and G.~Wendin:
Phys. Rev. B \textbf{72}, 014501 (2005); E.~V.~Bezuglyi,
A.~S.~Vasenko, E.~N.~Bratus, V.~S.~Shumeiko, and G.~Wendin:
\textit{ibid.} \textbf{73}, 220506(R) (2006)

\bibitem{Dynes}
J. P. Pekola, V. F. Maisi, S. Kafanov, N. Chekurov, A. Kemppinen, Yu. A. Pashkin, O.-P. Saira, M. Mottonen, and J. S. Tsai:
Phys. Rev. Lett. \textbf{105}, 026803 (2010)

\bibitem{LOnoneq}
A.~I.~Larkin and Yu.~N.~Ovchinnikov: in {\it Nonequilibrium Superconductivity},
edited by D.~N.~Langenberg and A.~I.~Lar\-kin (Elsevier, Amsterdam, 1986)

\bibitem{Belzig}
W.~Belzig, F.~K.~Wilhelm, C.~Bruder, G.~Sch\"{o}n, and A.~D.~Za\-i\-kin:
Superlatt. Microstruct. {\bf 25}, 1251 (1999)

\bibitem{VZK}
A.~F.~Volkov, A.~V.~Zaitsev, and T.~M.~Klapwijk: Physica C {\bf 210}, 21
(1993)

\bibitem{VBCH}
A.~S.~Vasenko, E.~V.~Bezuglyi, H.~Courtois, and F.~W.~J.~Hekking:
Phys. Rev. B {\bf 81}, 094513 (2010)

\bibitem{Bergeret}
F. S. Bergeret, A. F. Volkov, and K. B. Efetov: Phys. Rev. Lett. \textbf{86},
4096 (2001)

\bibitem{Ozaeta2}
A.  Ozaeta, A. S. Vasenko, F. W. J. Hekking   and F. S. Bergeret:
Phys. Rev. B \textbf{85}, 174518 (2012)

\bibitem{Demler}
E. A. Demler, G. B. Arnold and M. R. Beasley: Phys. Rev. B \textbf{55}, 15174 (1997)

\bibitem{Giazotto2006}
F.~Giazotto, T.~T.~Heikkil\"{a}, A.~Luukanen, A.~M.~Savin, and
J.~P.~Pekola: Rev. Mod. Phys. {\bf 78}, 217 (2006)

\bibitem{Pekola_A}
J. T. Muhonen, M. Meschke, and J. P. Pekola: Rep. Prog. Phys. \textbf{75}, 046501 (2012)

\bibitem{VH}
A.~S.~Vasenko and F.~W.~J.~Hekking: J. Low Temp. Phys. {\bf 154}, 221 (2009)

\bibitem{RCP}
S.~Rajauria, H.~Courtois, and B.~Pannetier: Phys. Rev. B {\bf 80}, 214521 (2009)

\bibitem{trap1}
J.~P.~Pekola, D.~V.~Anghel, T.~I.~Suppula, J.~K.~Suoknuuti, A.~J.\ Man\-ninen,
and M.~Manninen: Appl. Phys. Lett. {\bf 76}, 2782 (2000)

\bibitem{trap2}
D.~Golubev and A.~Vasenko: in {\it International Workshop on Superconducting
Nano-electronics Devices}, ed. by J.~Pekola, B.~Ruggiero, and P.~Silvestrini
(Kluwer Academic, Dordrecht 2002), p. 165; [arXiv:1204.2719]

\bibitem{Sukumar}
S.~Rajauria, P.~Gandit, T.~Fournier, F.~W.~J.~Hekking, B.~Pannetier, and
H.~Courtois: Phys. Rev. Lett. {\bf 100}, 207002 (2008)

\bibitem{Giazotto}
F.~Giazotto, F.~Taddei, R.~Fazio, and F.~Beltram: Appl. Phys.
Lett. \textbf{80}, 3784 (2002)

\end{thebibliography}
\end{document}